\documentclass[letter]{jpsj2} 
\title{
Anti-phase Modulation of Electron- and Hole-like States in Vortex Core of $\mathbf{Bi_2Sr_2CaCu_2O_x}$
Probed by Scanning Tunneling Spectroscopy}

\author{
Ken \textsc{Matsuba}$^{1}$, Shunsuke \textsc{Yoshizawa}$^{1}$, Yugo \textsc{Mochizuki}$^{1}$,
Takashi \textsc{Mochiku}$^{2}$,\\
Kazuto \textsc{Hirata}$^{2}$
and Nobuhiko \textsc{Nishida}$^{1}$ \thanks{E-mail address: nishida@ltp.phys.titech.ac.jp}}

\inst{$^{1}$Department of Physics, Tokyo Institute of Technology, Tokyo 152-8551, Japan\\
$^{2}$Superconducting Materials Center, National Institute for Materials Science, Tsukuba, Ibaraki 305-0047, Japan}

\abst{
In the vortex core of slightly overdoped \bi,
the electron-like and hole-like states have been found to exhibit spatial modulations in anti-phase with each other
along the Cu-O bonding direction.
Some kind of one-dimensionality has been observed in the vortex core,
and it is more clearly seen in differential conductance maps at lower biases below $\pm$9 mV.
}

\kword{vortex core, STS, STM, \bi, high temperature superconductor}

\begin{document}
\newcommand{\bi}{$\mathrm{Bi_2Sr_2CaCu_2O_x}$}
\newcommand{\nser}{$N_s(E, \mib{r})$}
\newcommand{\nspmnr}{$N_s(\mathrm{\pm 9meV}, \mib{r})$}
\newcommand{\nsnr}{$N_s(\mathrm{+9meV, \mib{r}})$ }
\newcommand{\didv}{$\mathrm{d}I/\mathrm{d}V(V, {\mib r})$}
\newcommand{\didvpmn}{$\mathrm{d}I/\mathrm{d}V(\pm\mathrm{9~mV}, {\mib r})$}
\newcommand{\didvn}{$\mathrm{d}I/\mathrm{d}V(+\mathrm{9~mV}, {\mib r})$}
\newcommand{\didvmn}{$\mathrm{d}I/\mathrm{d}V(-\mathrm{9~mV}, {\mib r})$}

\maketitle
The electronic states of vortex cores in type-II superconductors 
are closely related with the nature of the superconductivity.
As superconductivity is suppressed near the vortex, in BCS superconductors,
quasi-particles are confined in the vortex core 
by a pair potential $\Delta(r)$ reflecting the orbital symmetry of a Cooper pair, or 
in highly correlated electron system superconductors
other ordered states competing with superconductivity may appear. 
As scanning tunneling microscopy and spectroscopy (STM/STS) are able 
to measure the quasi-particle density of states in atomic length scale, 
microscopic STM/STS studies of vortex cores will give us 
valuable information to understand superconductivity.
In BCS superconductors, the vortex core bound states 
have been detected in 2H-NbSe$_2$ \cite{Hess} and YNi$_2$B$_2$C \cite{Nishimori}
and can be interpreted in terms of the Bogoliubov-de Gennes equation.
In high-T$_c$ cuprate superconductors,
vortex core states have been detected
in YBa$_2$Cu$_3$O$_7$ \cite{Maggio} and \bi \cite{Pan, Matsuba} by STM/STS, 
but they cannot be interpreted in terms of 
BCS $d_{x^{2}-y^{2}}$-superconductors. 
In \bi, it has been found that the vortex core state
exhibits a spatial modulation of STM conductance with a period of about 4$a_0$ \cite{Hoffman, Matsuba}
and the modulation is nondispersive \cite{Levy}.
Other competing orders might occur in the vortex core. 
Even in a zero magnetic field, STM/STS have revealed the tendency 
to exhibit a charge order in high-T$_c$ superconductors: 
Above $T_{c}$, the pseudo-gap state in slightly underdoped \bi~has been found
to exhibit a spatial modulation with an incommensurate periodicity ($\sim$4.7$a_{0}$) \cite{Vershinin}.
In very underdoped and insulating regions,
spatial modulations have been observed:
an incommensurate one ($\sim$4.5$a_{0}$) in \bi \cite{McElroy}
and a commensurate one (4$a_0$) in Ca$_{2-x}$Na$_{x}$CuO$_{2}$Cl$_2$ \cite{Hanaguri}.
For the present it is not clear how the conductance modulations in different situations are related with one another.
In this paper, we report high-resolution STS measurements of vortex cores
in slightly overdoped \bi.
Anti-phase modulation of electron-like and hole-like core states has been discovered
and some kind of one-dimensionality has been found in the vortex core.

The STS measurements were carried out by using a laboratory-made scanning tunneling microscope 
which is able to be operated at temperatures down to 2.2 K in magnetic fields up to 17 T.
A  mechanically-sharpened PtIr tip was used.
A single crystal sample of \bi~was grown by the traveling solvent floating zone method.
It is slightly overdoped and the critical temperature is 86 K.
A clean surface was prepared by cleaving the sample at 4.2 K in a He atmosphere.
We have measured a spatial distribution of the superconducting energy gap $\Delta_p$ by STS;
the average value was 37.8 meV and the variance was 4.8 meV.
A degree of disorder in ref. [11] was 0.126.

\begin{figure}[!p]
\begin{center}
\includegraphics[width=7.0cm]{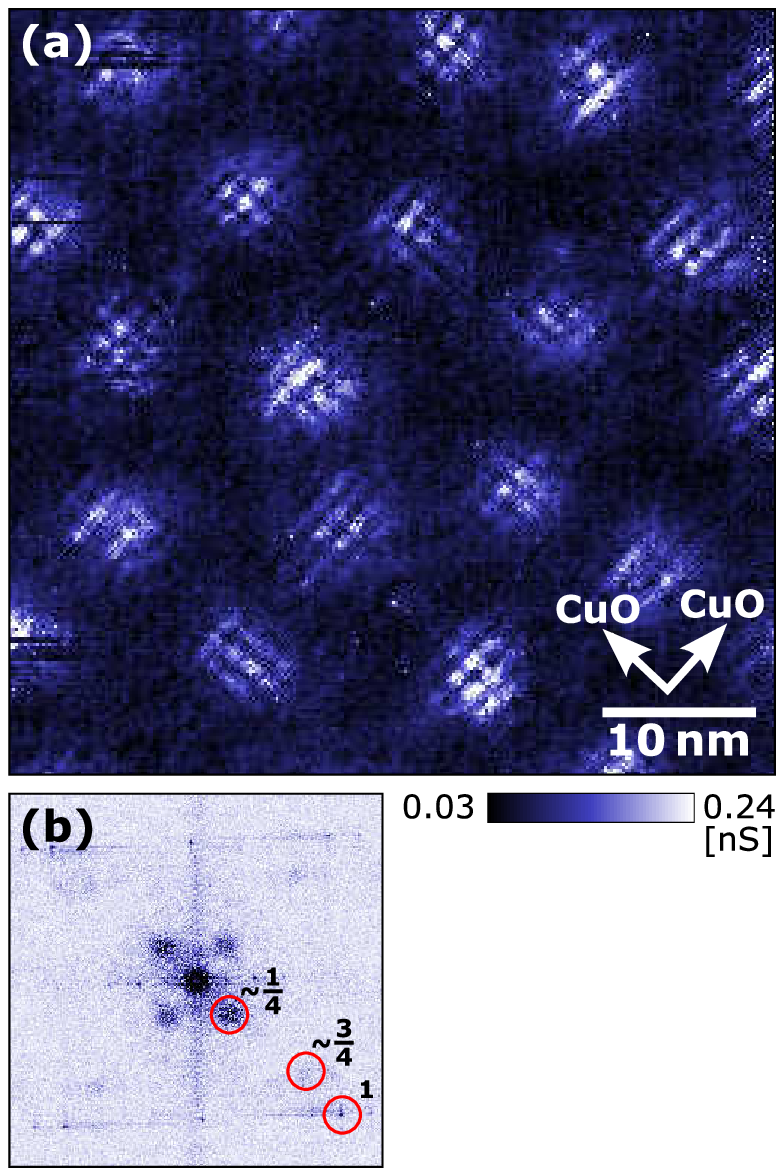}
\end{center}
\caption{(Color Online)
(a) Vortex cores of \bi~observed at 4.2 K in 14.5 T
in a region of 50 nm $\times$ 50 nm by scanning tunneling spectroscopy.
Two arrows indicate the Cu-O boding directions determined by the STM atomic image.
A periodic structure is observed in each core.
(b) The Fourier transform of (a). Three kinds of spots are seen, labeled as 1,  $\sim$1/4 and $\sim$3/4.}
\label{f1}
\end{figure}

We have observed vortex cores at 4.2 K  in 14.5 T in five regions of 50 nm $\times$ 50 nm
on the cleaved surface of \bi~by STS;
a typical result is shown in Fig. \ref{f1} (a).
The data taking procedure is as follows:
an STM image of 256 $\times$ 256 points is obtained
by scanning the tip on the sample surface at a tunneling current $I$ of 0.1 nA and a sample bias $V$ of +191 mV.
Simultaneously, a differential conductance \didv~is measured
by using a lock-in amplifier with a modulation bias of 1 $\mathrm{ mV_{rms}}$.
Previously we reported that vortex core states are situated at $\pm$9 meV 
in the energy gap \cite{Matsuba}.
By mapping \didvn~with a color scale, the core can be imaged
as shown in Fig. \ref{f1} (a). 
The vortices have been clearly imaged as white regions.
A periodic structure along Cu-O bonding directions has been observed in each core.
This pattern was called `checkerboard' in ref. [6].
We would rather call this a `tile and joint' pattern, 
as will be explained later. 
The Fourier transform is shown in Fig. \ref{f1} (b).
Three kinds of important spots are seen, as indicated by circles:
the spots $(\pm \frac{\mathrm 2\pi}{\lambda },0), (0,\pm \frac{\mathrm 2\pi}{\lambda})$
with $\lambda=a_{0}$
are due to Bi atoms of the BiO surface
and the other two $(\pm \frac{\mathrm 2\pi}{\lambda },0), 
(0,\pm \frac{\mathrm 2\pi}{\lambda})$
with $\lambda\sim \mathrm{4}a_{0}$ and $\lambda\sim\frac{4}{3}a_{0}$
are from the vortex core. 
The average periods of the periodic structure are (4.3 $\pm$ 0.2)$a_0$ and (1.34 $\pm$ 0.02)$a_0$.

\begin{figure*}[!p]
\begin{minipage}{\textwidth}
\begin{center}
\includegraphics[width=\textwidth]{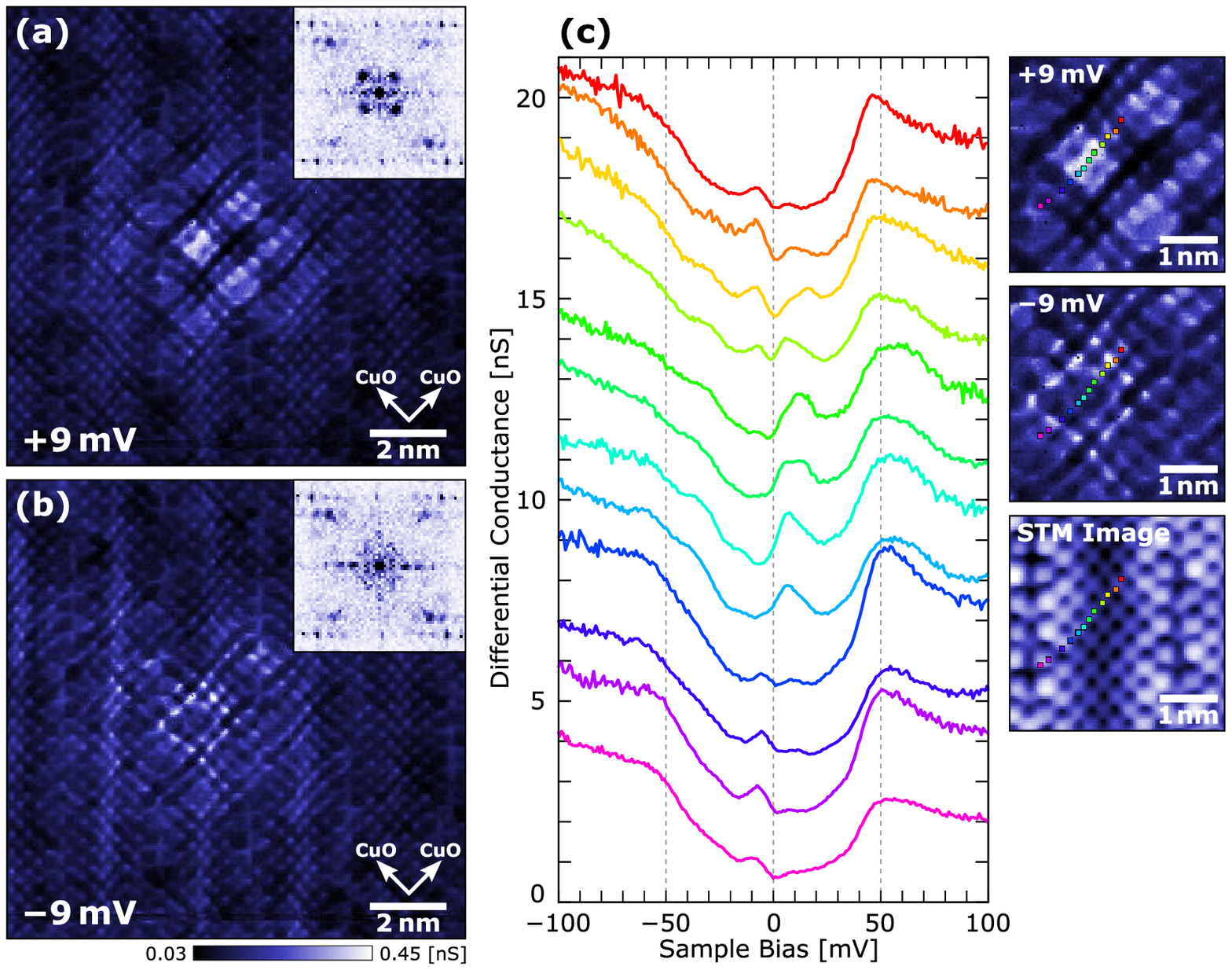}
\end{center}
\caption{
(a) and (b):
One vortex core measured at 4.2 K in 14.5 T by STS at 256 $\times$ 256 points
in a region of 12.0 nm $\times$ 12.0 nm. (a) the d$I$/d$V$ map at +9 mV and (b) at $-$9 mV;
the two d$I$/d$V$ images were simultaneously obtained together with the atomic image.
The Fourier transforms are shown in the insets.
(c):
Tunneling spectra were measured in the vortex core
by using a lock-in amplifier with a modulation bias of 1 $\mathrm{mV_{rms}}$,
after the tip-sample distance was adjusted at $I$ = 0.5 nA and $V$ = +191 mV.
The spectra are offset by 1.5 nS for clarity.
The measured positions are shown with square marks in the \didvpmn~maps and the STM atomic image;
the colors indicate the corresponding spectra.
The electron-like and hole-like core states are found to exhibit anti-phase spatial variation in intensity.
}
\label{f2}
\end{minipage}
\end{figure*}

We have scrutinized five vortices 
at 4.2 K in 14.5 T by STM/STS with a spatial resolution of 0.47 \r{A}.
The typical results for one vortex are shown in Fig. \ref{f2}:
(a) a d$I$/d$V$ image of the electron-like core state at +9 mV and (b) that of the hole-like core state at $-$9 mV. 
The modulated pattern of the vortex core at $\pm$9 mV looks like `tile and joint' rather than `checkerboard'.
Looking at Figs. \ref{f2} (a) and (b) carefully,
you will notice that the bright positions of \didvmn~coincide with
the dark positions of \didvn.
It means that in the vortex core the electron-like and hole-like states
exhibit anti-phase spatial variation in intensity.
The vortex core has been more clearly imaged at the positive bias than at the negative one,
as is reported before \cite{Matsuba}.
Outside the vortex, the in-phase variations of d$I$/d$V$ have been observed
at both positive and negative biases.
The atoms and the superstructure have been observed. 
The Fourier transforms of one vortex image in Figs. \ref{f2} (a) and (b) are shown in the insets.
 At +9 mV, the spots of $\sim$1/4 and $\sim$3/4 have been recognized.
The $\sim$1/4 spots have been found to be two-fold symmetric.
This will be discussed later.
At $-$9 mV, the $\sim$3/4 spots are observed clearly,
but those of $\sim$1/4 are found to be very weak, compared to at the positive bias.

Tunneling spectra have been  measured 
in the vortex core along the Cu-O bonding direction
with spacings of about 2 \r{A} (Fig. \ref{f2} (c)):
The measured positions are shown with square marks in \didvpmn~maps 
and the atomic STM image; the colors indicate the corresponding spectra.
The vortex core states are recognized as peaks at about $\pm$9 mV 
inside the energy gap.
From the top, in the first to third spectra, the peak at a negative bias is
more outstanding than that at a positive bias.
In the next fourth to eighth spectra, a reverse situation is seen;
the intensity at a positive bias is higher than the other.
In the last four spectra, the situation has been reversed again.
The anti-phase spatial variation of electron-like and hole-like core states has been confirmed.
The $\sim$3/4 modulation has been found to be in-phase,
because $\sim$3/4 peaks vanish in the Fourier transform
of the \didvn~normalized by \didvmn.
The results of the other four vortices investigated in the same manner 
are summarized as follows:
Three vortices have exhibited the same anti-phase variation.
In the other vortex where the insulating tunneling spectra have been 
observed around the center, the tile and joint  pattern was not observed 
and similar d$I$/d$V$ images were obtained at positive and negative biases;
the core states are found to be much influenced by defects.
We have concluded that the anti-phase correlation is intrinsic 
to a vortex core in \bi.
In the tunneling spectra of Fig. \ref{f2} (c),
it is noticed that the peak energy of the core states fluctuates 
from site to site over a length scale of a few angstroms.
We speculate that the peak energy might be influenced
by the excess oxygen atoms found by McElroy {\itshape et al.} \cite{McElroy2}.

\begin{figure}[!p]
\begin{center}
\includegraphics[width=6.7cm]{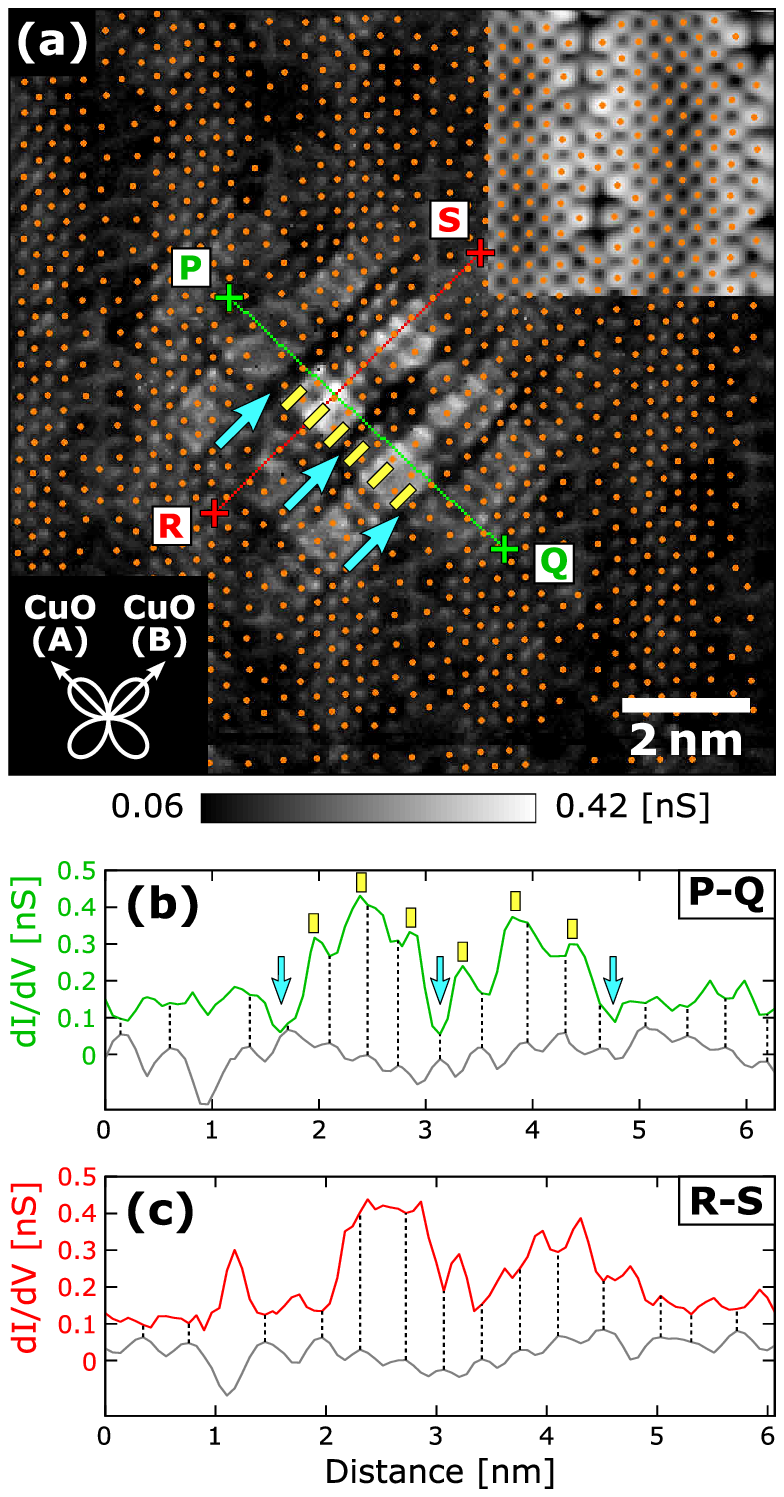}
\end{center}
\caption{
(a) One vortex core imaged by mapping \didvn~with a gray scale with atomic positions marked by orange dots.
The atomic image is shown in the upper-right inset and
the Cu-O bonding directions in the lower-left.
Three dark lines indicated by blue arrows are observed.
Three peaks are seen between the dark lines, as shown by yellow bars.
(b) and (c):
Line profiles of \didvn~along PQ (a green line) (b)  and RS (a red line) (c).
The topographic profiles of atoms are plotted by gray lines and
atom positions are indicated by broken lines.
}
\label{f3}
\end{figure}

In Fig. \ref{f2} (a), the one-dimensional nature will be noticed in the tile and joint pattern of the vortex;
the joints along one Cu-O direction are more clearly seen than those along the other. 
In order to know more clearly the situation,
the same image is shown in Fig. \ref{f3} (a) with Bi atom positions marked by orange dots;
the STM atomic image is shown in the upper right corner.
The d$I$/d$V$ line profiles along the green and red lines are shown in Figs. \ref{f3} (b) and (c) with the atomic profiles.
As indicated by blue arrows, three dark lines have been observed 
along the atomic rows: the spacing is exactly 4$a_0$.
The tile and joint pattern is found to have commensurate modulation 
along the CuO(A) direction.
Three peaks of d$I$/d$V$ are observed between the dark lines, 
as shown by yellow bars.
This fine structure corresponds to the $\sim$3/4 spot in the Fourier transform.
Along the CuO(B) direction, as it is difficult to get a precise period from the real space image,
the modulation period is determined to be 4.3$a_0$ from the Fourier transform (Fig. \ref{f2} (a)), 
indicating incommensurate variation along the CuO(B) direction.
It should be noted that different commensurability is observed 
in the two Cu-O bonding directions.
In the other vortex investigated, the dark lines were observed 
along the CuO(A) direction.
In the vortex core, two Cu-O bonding directions have become non-equivalent.
In fact, looking at Fig. \ref{f1} (a) carefully, you can see
some directionality along different Cu-O bonding directions from vortex to vortex.
A similar directional feature has been observed at a negative bias of $-$9 mV.

\begin{figure}[!p]
\begin{center}
\includegraphics[width=7.0cm]{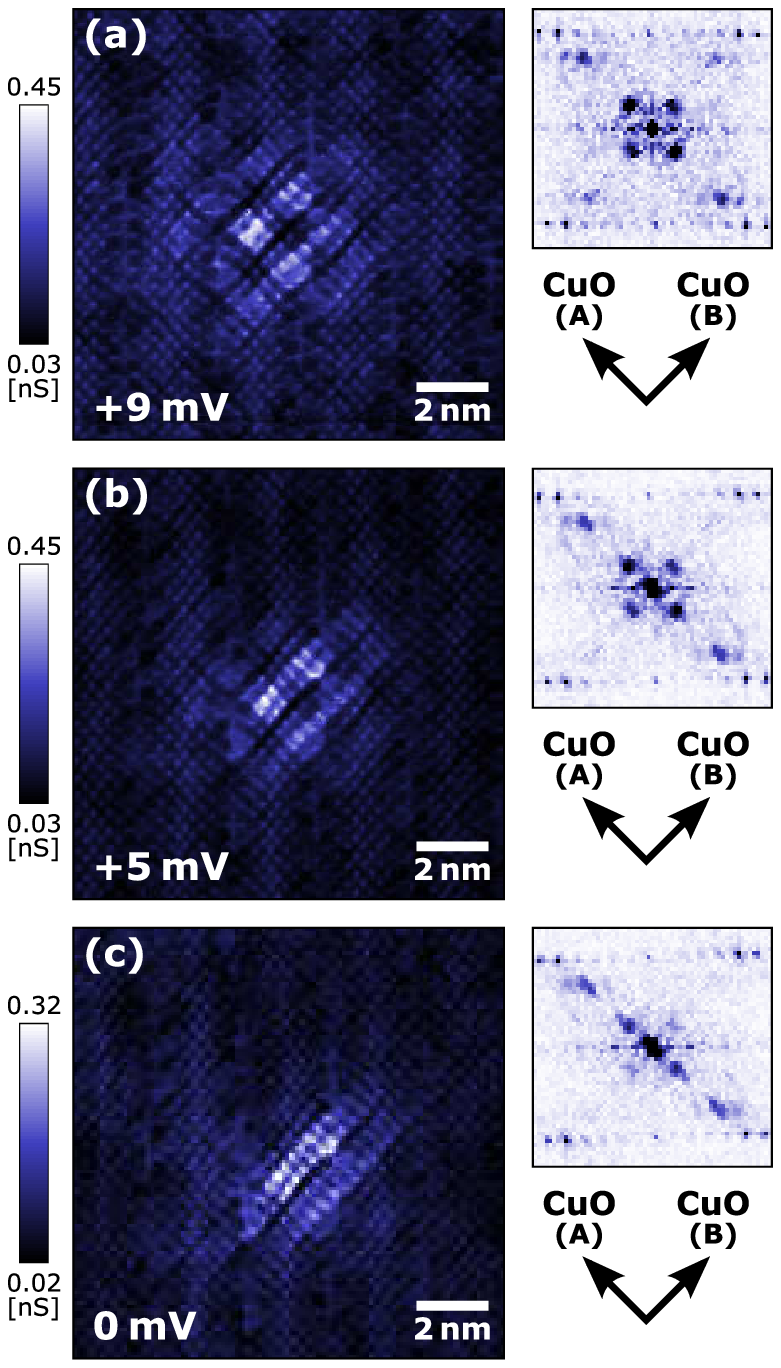}
\end{center}
\caption{
(Color online)
The same vortex core is imaged by mapping d$I$/d$V$ at (a) +9 mV, (b) +5 mV and (c) 0 mV.
The Fourier transform images are shown on the right side.
}
\label{f4}
\end{figure}

At different biases lower or higher than +9 mV, the same vortex core was imaged
by mapping d$I$/d$V$; 
the results at 0 and +5 mV are shown in Figs. \ref{f4} with that of +9 mV;
the Fourier transforms are also shown.
With decreasing a bias down to +5 and 0 mV [Figs. \ref{f4} (b) and (c)],
the one-dimensional nature has become more outstanding;
at +5 mV [Fig. \ref{f4} (b)], the dark line along the CuO(B) direction 
between two tiles become clearer.
The $\sim$3/4 spots have become vague along the CuO(B) direction,
while the $\sim$1/4 spots are still observed in the two directions.
At 0 mV [Fig. \ref{f4} (c)], the $\sim$1/4 spots have become vague along the CuO(B) direction 
and the $\sim$3/4 spots have disappeared.
Similar features were observed at negative biases.
Thus, it has been found that 
the vortex core state exhibits a two-fold symmetry at low energy 
rather than a four-fold one.
The origin of these features is unknown,
but the results will give valuable information to understand the conductance modulation in the vortex core.

The results at higher biases are described briefly:
the tile and joint pattern became unclear at $\pm$17 mV and it was invisible at $\pm$20 mV.
In the center region of the vortex core, only low energy states are present 
and the states  above $\pm$20 mV are not present.

In order to explain the electronic states of a vortex core in high temperature cuprate superconductors,
many theoretical proposals have been made.  
The theories \cite{Wang, Franz, Yasui} based on the $d_{x^{2}-y^{2}}$ BCS superconductor
have failed in explaining the experimental findings of vortex core states,
especially those revealed by STM/STS experiments.
When the superconductivity is weakened, 
similar modulated patterns were observed
in a zero magnetic field \cite{Vershinin, Hanaguri, McElroy, Machida}.
These ordered states may appear in the vortex core.
Vershinin {\itshape et al.} found a modulation in \bi~above $T_c$, that is,
in the pseudo-gap state \cite{Vershinin}.
The modulation is observed inside an energy gap ($\left|E\right|$ $<$ $\sim$40 meV).
The energy of the vortex core states, which we have observed 
with spatial modulation, is restricted to $\left|E\right|$ $<$ $\sim$20 meV;
at high temperatures, they are smeared out by thermal fluctuations.
In order to relate the core states with a pseudo-gap state,
we need more theoretical considerations, though it was proposed that
the core states correspond to the pseudo-gap state at low temperatures \cite{Renner, Levy}.
Hanaguri {\itshape et al.} and McElroy {\itshape et al.} found spatial modulations in very underdoped cuprates:
$\mathrm{Ca_{2-x}Na_{x}CuO_{2}Cl_{2}}$ \cite{Hanaguri} and \bi \cite{McElroy}.
They are observed at $\left|E\right|$ $<$ 100 meV in $\mathrm{Ca_{2-x}Na_{x}CuO_{2}Cl_{2}}$
and at 65 $<$ $\left|E\right|$ $<$ $\sim$100 meV in \bi.
These electronic states seem to be different from the core states,
as the energy range where modulations have been observed is very different.
Machida {\itshape et al.} found a conductance modulation at 0 $<$ $E$ $<$ 15 meV
in a single $\mathrm{CuO_{2}}$ layer cuprate
$\mathrm{Bi_{2}Sr_{1.6}La_{0.4}CuO_{6+\delta}}$ \cite{Machida} ($T_c$ $\sim$ 34 K).
The tunneling spectra are very similar to those in the vortex core. 
The close similarity is interesting.
In order to compare the above-mentioned states with the vortex core states,
it is necessary to study the phase relation of the modulation in electron-like and hole-like states.

In summary, in the vortex core of slightly overdoped \bi , 
it has been found that the electron-like and hole-like states exhibit
the anti-phase spatial modulation along the Cu-O bonding directions and 
that at low energy below $\pm$9 meV, the two Cu-O bonding directions are non-equivalent
and the vortex core states have a two-fold symmetry rather than a four-fold one of outside.
The modulation period along one Cu-O direction is 4.3$a_0$ (incommensurate) and
that along the other is 4.0$a_0$ (commensurate).
Some ordered state must be realized in the vortex core at low energy.

We acknowledge T. Okumura, Y. Kiuchi, Y. Kikuchi, T. Ogawa and N. Kosugi 
for data taking and analyses in the early stage of the experiment.
This work was supported in part by a Grant-in-Aid for Scientific Research
of the Ministry of Education, Culture, Sports, Science and Technology (grant No.14204035)
and a 21st Century COE Program at Tokyo Tech ``Nanometer-Scale Quantum Physics."
One of the authors (K.M.) acknowledges Research Fellowships of the Japan Society
for the Promotion of Science for Young Scientists.

\begin{thebibliography}{99} 
\bibitem{Hess} H. F. Hess, R. B. Robinson, R. C. Dynes, J. M. Valles, Jr. and J. V. Waszczak: Phys. Rev. Lett. \textbf{62} (1989) 214.
\bibitem{Nishimori}H. Nishimori, K. Uchiyama, S. Kaneko, A. Tokura, H. Takeya, K. Hirata and N. Nishida: J. Phys. Soc. Jpn. \textbf{73} (2004) 3247.
\bibitem{Maggio} I. Maggio-Aprile, Ch. Renner, A. Erb, E. Walker and \O. Fischer: Phys. Rev. Lett.
\textbf{75} (1995) 2754.
\bibitem{Pan} S. H. Pan, E. W. Hudson, A. K. Gupta, K.-W. Ng, H. Eisaki, S. Uchida and J. C. Davis: Phys. Rev. Lett. \textbf{85} (2000) 1536.
\bibitem{Matsuba} K. Matsuba, H. Sakata, N. Kosugi, H. Nishimori and N. Nishida: J. Phys. Soc. Jpn. \textbf{72} (2003) 2153.
\bibitem{Hoffman} J. E. Hoffman, E. W. Hudson, K. M. Lang, V. Madhavan, H. Eisaki, S. Uchida and J. C. Davis: Science \textbf{295} (2002) 466.
\bibitem{Levy} G. Levy, M. Kugler, A. A. Manuel, and \O. Fischer: Phys. Rev. Lett. \textbf{95} (2005) 257005.
\bibitem{Vershinin} M. Vershinin, S. Misra, S. Ono, Y. Abe, Y. Ando and A. Yazdani: Science \textbf{303} (2004) 1995.
\bibitem{McElroy} K. McElroy, D.-H. Lee, J. E. Hoffman, K. M. Lang, J. Lee, E. W. Hudson, H. Eisaki, S. Uchida and J. C. Davis: Phys. Rev. Lett. \textbf{94} (2005) 197005.
\bibitem{Hanaguri} T. Hanaguri, C. Lupien, Y. Kohsaka, D.-H. Lee, M. Azuma, M. Takano, H. Takagi and J. C. Davis: Nature \textbf{430} (2004) 1001.
\bibitem{Matsuba2} K. Matsuba, H. Sakata, T. Mochiku, K. Hirata and N. Nishida: Physica C \textbf{388-389} (2003) 281.
\bibitem{McElroy2} K. McElroy, J. Lee, J. A. Slezak, D.-H. Lee, H. Eisaki, S. Uchida and J. C. Davis: Science \textbf{309} (2005) 1048.
\bibitem{Wang} Y. Wang and A. H. MacDonald: Phys. Rev. B \textbf{52} (1995) R3876.
\bibitem{Franz} M. Franz and Z. Te\v{s}anovi\'{c}: Phys. Rev. Lett. \textbf{80} (1998) 4763.
\bibitem{Yasui} K. Yasui and T. Kita: Phys. Rev. Lett.  \textbf{83} (1999) 4168.
\bibitem{Renner} Ch. Renner, B. Revaz, K. Kadowaki, I. Maggio-Aprile and \O. Fischer: Phys. Rev. Lett. \textbf{80} (1998) 3606.
\bibitem{Machida} T. Machida, Y. Kamijo, K. Harada, T. Noguchi, R. Saito, T. Kato and H. Sakata: J. Phys. Soc. Jpn. \textbf{75} (2006) 083708.
\end{thebibliography}
\end{document}